\documentclass[aps,prl,preprint,groupedaddress]{revtex4}

\usepackage{graphicx}
\begin{document}
\title{Helium $2\,^3{\rm S}$ - $2\,^1{\rm S}$ metrology at $1.557\,\mu{\rm m}$}
\author{K.\ A.\ H.\ van Leeuwen$^{1,2}$}
\author{W.\ Vassen$^1$}
\email{w.vassen@few.vu.nl}
\affiliation{$^1$Laser Centre Vrije Universiteit, De Boelelaan 1081, 1081 HV Amsterdam, The Netherlands}
\affiliation{$^2$Department of Applied Physics, Eindhoven University of Technology, P.O. Box 513, 5600 MB Eindhoven, The Netherlands}




\begin{abstract}
An experiment is proposed to excite the `forbidden' $1s2s\,^3{\rm
S}_1$ - $1s2s\,^1{\rm S}_0$ magnetic dipole (M1) transition at
$1.557\,\mu{\rm m}$ in a collimated and slow atomic beam of
metastable helium atoms. It is demonstrated that an excitation
rate of $5000\,{\rm s}^{-1}$ can be realised with the beam of a
$2\,{\rm W}$ narrowband telecom fiber laser intersecting the
atomic beam perpendicularly. A Doppler-limited sub-MHz 
spectroscopic linewidth is anticipated. Doppler-free
excitation of 2\% of trapped and cooled atoms may be realised in 
a one-dimensional optical lattice geometry, using the 2 W laser both
for trapping and spectroscopy. 
The very small (8 Hz) natural
linewidth of this transition presents an opportunity for accurate
tests
of atomic structure
calculations of the helium atom. A measurement of the
$^3$He - $^4$He isotope shift allows for accurate determination of
the difference in nuclear charge radius of both isotopes.
\end{abstract}

\maketitle

\section{Introduction}
Measurements of level energies of low-lying states in helium
provide very sensitive tests of basic theory of atomic structure.
Although helium is a two-electron system, energy levels of the
nonrelativistic helium atom can be calculated with a precision
that is, for all practical purposes, as good as for
nonrelativistic hydrogen. Relativistic corrections to these
energies can be calculated in a power series of the fine structure
constant $\alpha$ and have been calculated up to $O(\alpha^2)$.
Effects of the finite nuclear mass can be included in a power
series of the mass ratio $\mu/M$, where $\mu$ is the reduced
electron mass and $M$ the nuclear mass. QED corrections, both
hydrogenic and electron-electron terms, up to $O(\alpha^3)$ have
been calculated as well. Effects of the finite nuclear size can be
incorporated straightforwardly~\cite{Drake98, Drake04}. Present
day theory aims at calculating higher-order corrections (in
$\alpha$, $\mu/M$) and cross terms (such as relativistic recoil).
The most difficult terms to date are the relativistic and QED
terms of $O(\alpha^4)$ and higher. Higher-order corrections are
largest for low-lying S-states and therefore the most sensitive
tests of atomic structure calculations can be performed for the
$1\,^1{\rm S}_0$ ground state and the metastable states $2\,^1{\rm
S}_0$ (lifetime 20 ms) and $2\,^3{\rm S}_1$ (lifetime $
8000\,{\rm s}$).

Present-day laser spectroscopy on the low-lying S-states has
several disadvantages. To extract an experimental value for the
ionisation energy of these states the transition frequency 
to a high-lying state has to
be accurately measured and one has to rely on theoretical values
of the ionisation energy of the upper state in the transition.
Here the lifetime of the upper state and line shifts due to stray
electric fields and laser power are limiting factors. For the
ground state, excitation is difficult: one photon of $58\,{\rm nm}$
is required to excite the $2\,^1{\rm P}_1$ state \cite{Eikema97}
or two photons of $120\,{\rm nm}$ to excite the $2\,^1{\rm S}_0$
state \cite{Bergeson98}.
For the $2\,^1{\rm S}_0$ and $2\,^3{\rm S}_1$ metastable states CW
laser light is used to excite with one photon the $2\,^3{\rm P}$
\cite{CancioPastor04}, $3\,^3{\rm P}$ \cite{Pavone94, Mueller05} and ${\rm
n}\,^1{\rm P}$~\cite{Sansonetti92} states. Two-photon spectroscopy
is applied to excite the $3\,^3{\rm D}$ state \cite{Dorrer97} and
${\rm n}\,^1{\rm D}$ states \cite{Lichten91}.

The most accurate transition frequency measurement to date is for
the $2\,^3{\rm S}_1 \rightarrow 2\,^3{\rm P}_0$
transition~\cite{CancioPastor04}, with an absolute accuracy of
$4\,{\rm kHz}$. This measurement, however, does not provide an
accurate measurement of the $2\,^3{\rm S}_1$ ionisation energy as
the $2\,^3{\rm P}_0$ ionisation energy is not known well enough.
Moreover, recent measurements of the fine structure splitting of
the $2\,^3{\rm P}$ state have shown that experiment and theory do
not agree at the $10\,{\rm kHz}$ level~\cite{Pachucki06a}.
This very recent finding is considered an outstanding problem of
bound state QED and asks for independent measurements on other
transitions with similar accuracy. Experimental accuracies for the
ionisation energy of the $2\,^1{\rm S}_0$ and $2\,^3{\rm S}_1$
states, deduced from measurements to highly excited states and
relying on the theoretical calculations of the ionisation energies
of these states, are $150\,{\rm kHz}$ and $60\,{\rm kHz}$
respectively. These values agree with but are more accurate than
present-day theory for the 2 $^1$S$_0$ and 2 $^3$S$_1$ ionisation
energy, which is accurate to 5 MHz \cite{Morton06a} and 1 MHz
\cite{Pachucki00} respectively. The theoretical accuracies
represent the estimated magnitude of uncalculated higher-order
terms in QED calculations.

QED shifts largely cancel when identical transitions in different
isotopes are studied. Measuring the isotope shift,
the main theoretical inaccuracy is in the difference
in the rms charge radius of the nuclei \cite{Drake05, Morton06b}. As the
charge radius of the $^4$He nucleus is the most accurately known
of all nuclei (including the proton) isotope shifts
measure the charge radius of the other isotope involved. In this
way accurate determination of the 2 $^3$S$_1$ - 2 $^3$P transition
isotope shift have allowed accurate measurements of the charge
radius of $^3$He \cite{Shiner95, Morton06b} and the unstable isotope $^6$He
\cite{Wang04}, challenging nuclear physics calculations.
It has to be noted, however, that the measurement of the $^4$He nuclear 
radius has sofar not been reproduced \cite{Morton06b}. These measurements
therefore primarily measure differences in charge radius.


Elaborating on an idea of Baklanov and Denisov~\cite{Baklanov97}
we propose direct laser excitation of the
$2\,^3{\rm S}_1 \rightarrow 2\,^1{\rm S}_0$ transition in a slow atomic beam or
in an optical lattice.
This transition has the advantage of an intrinsically narrow natural linewidth of 8 Hz and a
wavelength of $1.557\,\mu$m.
Also, as the transition connects two states of the same $1s2s$
configuration, the theoretical error in the transition frequency
may be smaller than the error in the ionisation energy of the
individual states \cite{Pachucki06b}. The main disadvantage is that the transition is
extremely weak; the Einstein A-coefficient for this magnetic
dipole (M1) transition is 14 orders of magnitude smaller than for
the electric dipole (E1) $2\,^3{\rm S}_1 \rightarrow 2\,^3{\rm P}$
transition. 
In this paper we show that with 2 W of a narrowband fiber laser
at $1.557\,\mu$m we can excite more than 1 in $10^7$ atoms in
an atomic beam experiment or more than 1\% of atoms trapped
in a one-dimensional standing light wave. Present-day sources
of metastable helium atoms~\cite{Baldwin05} easily provide 
sufficient atoms in either an atomic beam or in a trap to 
observe the transition.

\section{Experimental feasibility}
The magnetic dipole transition between the metastable $2\,^3{\rm
S}_1$ state and the $1\,^1{\rm S}_0$ ground state determines the
$\approx 8000\,{\rm s}$ lifetime of the $2\,^3{\rm S}_1$
state. The experimental and theoretical values of the
rate constant for this transition,
$1.10(33)\times10^{-4}$ s$^{-1}$ \cite{Woodworth75} and 
$1.272\times10^{-4}\,{\rm s}^{-1}$
\cite{Drake71, Lach01} respectively, agree very well. For the electronically
similar transition from $2\,^3{\rm S}_1$ to $2\,^1{\rm S}_0$, rate constants
of $1.5 \times 10^{-7}\,{\rm s}^{-1}$~\cite{Lin77} and $6.1 \times
10^{-8}\,{\rm s}^{-1}$~\cite{Baklanov97} have been published. In
this letter we will use the value $9.1\times10^{-8}$ s$^{-1}$,
obtained by Pachucki~\cite{Pachucki06b} applying the same formalism
as used for the calculation of the $2\,^3{\rm S}_1$ to $1\,^1{\rm
S}_0$ transition rate \cite{Lach01}. This value is assumed to be
accurate at the 1\% level.

In order to evaluate the feasibility of detecting the transition,
we need to choose an experimental configuration. Several
approaches can be considered~\cite{Baklanov05}: spectroscopy in a
discharge cell, on a thermal or laser-cooled beam, on a cold cloud
in a magnetic or optical trap, and ultimately spectroscopy on
atoms in an optical lattice. Here, we first consider a relatively
simple but promising setup: spectroscopy on a laser-cooled and
collimated beam as produced on a daily basis in groups working on
BEC in metastable helium~\cite{Baldwin05}.

\section{Bloch equations model}
In order to describe the excitation of the $2\,^3{\rm S}_1
\rightarrow 2\,^1{\rm S}_0$ magnetic dipole transition, we use a
simplified set of optical Bloch equations (OBE's).
\begin{figure}
\begin{center}
\includegraphics[width=120mm]{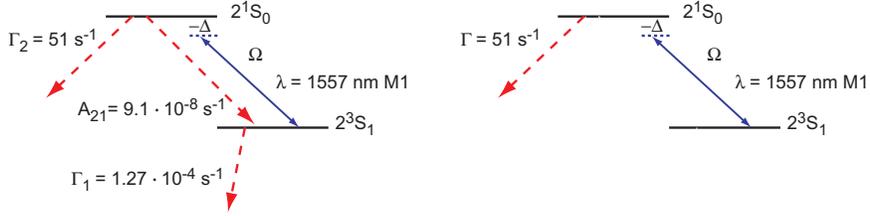}
\end{center}
 \caption{Relevant levels, transitions and decay rates (left) and simplified model (right)}
 \label{Fig:transition}
\end{figure}
The decay rate of the upper level is fully dominated by the
two-photon (2E1) decay to the ground state: $\Gamma_2=51\, {\rm
s}^{-1}$ \cite{Lin77}. Fig.~\ref{Fig:transition} summarizes the relevant decay
rates. The driven Rabi frequency of the transition (with atomic frequency $\omega_{21}$) is denoted by
$\Omega$, the detuning by $\Delta=\omega_{laser}-\omega_{21}$.

We model this transition by the following set of OBE's:

 \begin{eqnarray}
 \dot{\rho_{11}}=A_{21}\rho_{22}-\Gamma_1 \rho_{11}+i
      \frac{\Omega}{2}(\rho_{21}-\rho_{12})\\
 \dot{\rho_{22}}=-\Gamma_2 \rho_{22}-i
      \frac{\Omega}{2}(\rho_{21}-\rho_{12})\\
 \dot{\rho_{12}}=-\left(\frac{\Gamma_1+\Gamma_2}{2}+i \Delta\right)
      \rho_{12} + i \frac{\Omega}{2}(\rho_{22}-\rho_{11})\\
\dot{\rho_{21}}=-\left(\frac{\Gamma_1+\Gamma_2}{2}-i \Delta\right)
      \rho_{21} - i \frac{\Omega}{2}(\rho_{22}-\rho_{11})
 \end{eqnarray}

As the ground state is not included, this set of equations does
not have a steady state state solution except the trivial one
($\rho_{11}=\rho_{22}=\rho_{12}=\rho_{21}=0$).

When we simplify this set of equations by neglecting $A_{21}$ and
$\Gamma_1$ (which is certainly valid for times $t \ll 1/\Gamma_1 \approx 8000\,{\rm s}$),
it can be solved analytically.
Starting at $t=0$ with all atoms in the 2 $^3{\rm S}_1$ state
($\rho_{11}=1$), and denoting $\Gamma_2$ by $\Gamma$ from now on,
the result for the population $\rho_{22}$ of the 2 $^1{\rm S}_0$
state is:

\begin{equation}
\rho_{22}(t)= \frac{\Omega^2}{2\widetilde{\Omega}^2}\left[
\cosh{\left(
\sqrt{\frac{1}{2}[\widetilde{\Omega}
^2-\widehat{\Omega}^2+\frac{\Gamma^2}{4} ]
}\,\,t\right)}
-\cos{\left(\sqrt{\frac{1}{2}[\widetilde{\Omega}
^2+\widehat{\Omega}^2-\frac{\Gamma^2}{4}]
}\,\,t\right)}\right] e^{-\frac{\Gamma}{2}t},
\label{Eq:rho22}
\end{equation}
where $\widetilde{\Omega}^2 \equiv
\sqrt{(\Omega+\frac{\Gamma}{2})^2+\Delta^2}\sqrt{(\Omega-\frac{\Gamma}{2}
)^2+\Delta^2}$ and $\widehat{\Omega}^2 \equiv \Omega^2+\Delta^2$.

%

%

In this Letter, two limiting cases are studied, both valid for
short time ($t \ll 2/\Gamma=40$~ms): the weak excitation
limit ($1/t \gg \Omega \gg \Gamma/2$) and the strong excitation
limit ($\Omega \gg 1/t \gg \Gamma/2$). For the weak excitation
limit, Eq.~\ref{Eq:rho22} reduces to:
\begin{equation}
\rho_{22}(t)\approx \frac{\Omega^2}{\Delta^2}
\sin^2(\Delta\,\,t/2) \equiv
\left[\frac{\sin(\Delta\,\,t/2)}{\Delta\,\,t/2}\right]^2
\frac{\Omega^2}{4}t^2.
\label{Eq:weak}
\end{equation}

In the strong excitation limit, we can derive a simple expression for the upper
level population time-averaged over the oscillations of the cosine in Eq.~\ref{Eq:rho22}:
\begin{equation}
\langle \rho_{22} \rangle \approx
\frac{\Omega^2}{2(\Omega^2+\Delta^2)}
\label{Eq:strong}
\end{equation}

\section{Broadening effects}
Eqs.~\ref{Eq:weak} and \ref{Eq:strong} allow us to account effectively 
for both the finite bandwidth of the excitation laser and 
Doppler broadening effects. For this, we will assume
excitation by a purely inhomogeneously broadened light source,
i.e., the light is described by a monochromatic field with fixed
irradiance $I_0$ and a statistical probability distribution
$P(\omega)$ for the frequency. The distribution is assumed to be
Gaussian, centered at the transition frequency $\omega_{21}$ and
with an rms width $\Delta\omega_{rms}\gg 1/\tau$ with $\tau$ the
interaction time, i.e., the time at which the upper state
population is calculated. The Rabi frequency is given by
$\Omega^2=\frac{6 \pi\,c^2}{\hbar\omega_{21}^3}
A_{21}\langle\,J_1\,M_1\,1\,-q\,|J_2\,M_2\rangle ^2\,I_0$ for a
$|J_1, M_1\rangle\, \rightarrow |J_2,M_2\rangle$ transition
excited by light with polarisation $-q=-1,0,1$. With $J_1=1$ and
$J_2=0$, the non-zero squared Clebsch-Gordan coefficients all
equal $\frac{1}{3}$. Assuming the metastable (lower state) atoms
to be equally distributed over the three magnetic substates and
the polarisation to be pure but arbitrary, the total upper state
population can now be evaluated for both the weak and strong
excitation limits by integrating Eqs.~\ref{Eq:weak} and
\ref{Eq:strong}:
\begin{eqnarray}
\widetilde{\rho_{22}(\tau)}=\frac{\pi^2 c^2}{3 \hbar
\omega_{21}^3}A_{21} \frac{I_0}{\sqrt{2 \pi}\Delta\omega_{rms}}
\tau & \textrm{weak excitation limit} \label{Eq:rhoiweak}\\
\widetilde{\langle \rho_{22}\rangle}=\frac{\pi
c}{6}\sqrt{\frac{A_{21}I_0}{\hbar\omega_{21}^3}}
\frac{1}{\Delta\omega_{rms}} & \textrm{strong excitation limit}
\label{Eq:rhoistrong}
\end{eqnarray}

\section{Beam experiment}
Here we estimate the feasibility of an in-beam spectroscopic
experiment on the $1.557\,\mu{\rm m}$ He$^*$ $2\,^3{\rm S}_1
\rightarrow 2\,^1{\rm S}_0$ forbidden transition. The atomic beam is
Zeeman-slowed to $100\,{\rm m\,s}^{-1}$ and is transversely cooled
to an rms velocity spread of twice the Doppler limit for the
He$^*$ $2\,^3{\rm S}_1 \rightarrow 2\,^3{\rm P}_2$ cooling
transition ($\Delta v_{rms}=0.6\,\,{\rm ms}^{-1})$. The beam has
a diameter of $1\,{\rm cm}$ and an atom flux
$\Phi=10^{11}\,{\rm s}^{-1}$.
The transition is excited by the beam of a 2 Watt CW fiber laser
with a linewidth of $100\,{\rm kHz}$, intersecting the atomic beam
perpendicularly. For simplicity, we will assume a
flat-top square intensity profile of size $D_x \times D_y = 10
\times 10 \,{\rm mm}$.

In this experiment, the inverse of the interaction time
$1/\tau_{int}=10^4\,{\rm s^{-1}}$, the on-resonance Rabi frequency
of the transition $\Omega = \sqrt{\frac{2\pi
c^2}{\hbar\omega_{21}^3} A_{21} \frac{P}{D_x D_y}} = 74 \,{\rm
s}^{-1}$, and half the decay rate $\Gamma/2=25\,{\rm s^{-1}}$. We
are sufficiently in the weak excitation limit in this case to allow the use of
Eq.~\ref{Eq:rhoiweak}. The total upper state population now
evaluates to $\tilde{\rho}_{22}(\tau)= 4.6 \times 10^{-7}$. This
results in a flux of excited atoms
$\Phi_e=\tilde{\rho}_{22}(\tau)\Phi =4.6 \times 10^{3}\,{\rm
s}^{-1}$.

Using $1083\,{\rm nm}$ light resonant with the $2\,^3{\rm
S}_1\rightarrow 2\,^3{\rm P}_2$ transition, after the interaction
region the non-excited fraction of the atoms can be deflected by
simple radiation pressure. Using surface ionisation and an
electron multiplier on the non-deflected upper state atoms, in
principle all excited atoms ($2\,^1{\rm S}_0$) can then be detected.
$2\,^1{\rm S}_0$ atoms produced by the discharge beam source can be
very efficiently suppressed in the laser cooling stages used to
prepare the atomic beam. Thus, we can expect a workable signal.

\begin{figure}
\begin{center}
\includegraphics[width=150mm]{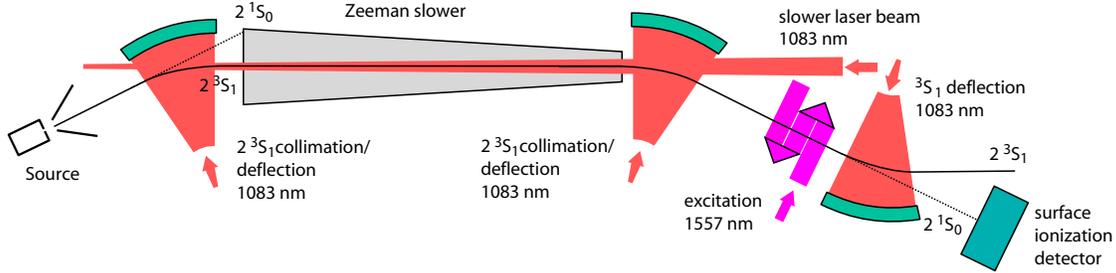}
\end{center}
 \caption{Schematic view of proposed setup.}
 \label{Fig:setup}
\end{figure}

The spectroscopic linewidth of $400\,{\rm kHz}$ is dominated by
the Doppler width. Decreasing the Doppler width
as well as reducing the laser
linewidth are the first steps towards decreasing the spectroscopic
linewidth. The limiting homogeneous linewidth is given by the
$10\,{\rm kHz}$ interaction time broadening. Applying multiple
crossings of laser-- and atomic beam using roof-top prisms will
not only increase the signal proportionally, but also decrease the
interaction time broadening. As a bonus, possible Doppler shifts
due to nonorthogonal excitation can be monitored and minimised in
this configuration. A schematic view of the proposed setup, with
three crossings depicted, is shown in Fig.~\ref{Fig:setup}.

\section{Experiment in an optical lattice}
The $1.557\,\mu{\rm m}$ wavelength of the forbidden transition can
also serve well to form a far off-resonance optical trap (FORT)
for the $2\,^3{\rm S}_1$ atoms. The dynamic polarisability at this
wavelength is fully dominated by the contribution of the $2\,^3{\rm
S}_1 \rightarrow 2\,^3{\rm P}_2$ transition that is resonant at
$1.083\,\mu{\rm m}$. This can be used in a next-generation
spectroscopy experiment. As the trapping potential is fully
insensitive to the wavelength on the scale of a spectroscopic scan
of the forbidden transition, a single laser can be simultaneously
used for trapping and spectroscopy. We propose using a cold
atomic cloud and transferring this cloud to a 
one-dimensional optical dipole trap formed by a single
retroreflected laser beam.

We assume $5 \times 10^7$ atoms in a
cloud with rms size $68\,\mu{\rm m}$ (radial) by $340\,\mu{\rm m}$
(axial) at a temperature of $5\,\mu{\rm K}$, as typically produced
along the route to BEC production \cite{Tychkov06}. The average density in this
cloud then equals $n_0=1.0 \times 10^{12}\,{\rm cm}^{-3}$.
We now assume a more stable 2 Watt fiber laser with a linewidth of
$10\,{\rm kHz}$.  The trapping/spectroscopy laser beam, directed
along the long axis of the cigar-shaped cloud, has a waist radius of
$136\,\mu{\rm m}$. This produces a trapping dipole potential with
a maximum depth of $24.4\,\mu{\rm K}$, to which the full cloud can
be transferred. Analysis of this trap indicates that roughly half
the atoms end up in the lowest vibrational state of the standing
wave ``micro-traps'' along the laser beam axis.
This will effectively cause a strong Doppler-free part in the
spectroscopic signal by Lamb-Dicke narrowing of the transition.
The Rabi frequency
$\Omega = 2\sqrt{\frac{2\pi c^2}{\hbar
\omega_{21}^3} A_{21} \frac{P}{D_x D_y}} = 6.28 \times 10^3\,{\rm
s}^{-1}$, where the extra factor of two is due to the fact that the
atoms are now trapped at the antinodes of the standing wave.
There is now no \emph{a priori} fixed interaction time. However,
as the Rabi frequency is high we can easily choose the
excitation to last a time $t$ satisfying the time-averaged strong
excitation limit ($\Omega \gg 1/t \gg \Gamma/2$): $t \gg 0.2$ ms.
Eq.~\ref{Eq:rhoistrong} then results in an averaged excited state
fraction $\widetilde{\langle
\rho_{22}\rangle}=\frac{1}{6}\sqrt{\frac{\pi}{2}}\frac{\Omega}{\Delta\omega_{rms}}=2.08
\times 10^{-2}$.
Given the number of atoms in the trap, every trapped sample will
lead to $10^6$ atoms in the excited state, which have to be detected. 
In an experiment, the
trapping/excitation laser can be set off-resonance for trapping,
and switched to the frequency of the forbidden transition to
excite the atoms to the $2\,^1{\rm S}_0$ state. We can then
detect the excited atoms by photoionisation after a few milliseconds of
excitation (adjusting the geometry of the ionisation laser beam to
ionise all $2\,^1{\rm S}_0$ atoms expelled from the trap). 

An alternative detection option is simply by measuring the increased Penning
ionisation that we expect when atoms are excited.
The excited $2\,^1{\rm S}_0$ atoms can also
decay through Penning ionisation in collisions with the
$2\,^3{\rm S}_1$ atoms. We assume the rate constant of this
process to be $10^{-10}\,{\rm cm}^3{\rm s}^{-1}$, i.e., on the
order of the rate constant for Penning ionisation of unpolarised
atoms in the $2\,^3{\rm S}_1$ state~\cite{Stas06}. This results in a decay rate
of $\Gamma_{PI}=100\,{\rm s}^{-1}$, leading to a homogeneous
broadening of the transition with this value.
However, as the dipole shift of the excited $2\,^1{\rm S}_0$ state
has opposite sign, the excited atoms are anti-trapped and escape
the cloud of $2\,^3{\rm S}_1$ atoms in $\tau_{esc} \approx
0.3\,{\rm ms}$. Then Penning ionisation stops,
effectively reducing the total $2\,^1{\rm S}_0$ ionisation rate by
a factor of 30 as compared to the case of perfect overlap.

As the Doppler width is effectively eliminated by Lamb-Dicke
narrowing, the linewidth is now determined by the laser
linewidth. As this width can be further reduced, ultimately the
limiting factor will be simply the upper level decay rate.

%
However, systematic shifts of the transition frequency will have
to be carefully considered and corrected for. The largest shift is
caused by the difference in dipole shift between the lower and
upper level of the transition and amounts to $2.0\,{\rm MHz}$ for
the chosen lattice parameters. In optical lattice experiments
aiming at optical frequency standards one selects a lattice laser
wavelength at which the Stark shift of the lower state equals the
Stark shift of the upper state (the `magic' wavelength). For
helium there is no practical wavelength available. The highest
magic wavelengths are around $410\,{\rm nm}$, where (accidentally)
the dipole shift itself is so small that no lattice is
feasible at reasonable laser power, and at $351\,{\rm nm}$, where 
the polarisability is 15 times smaller than at 1.557 $\mu$m and 
the sign is such that the atoms cannot be confined at antinodes. At 
1.557 $\mu$m, 
combining measurements of the transition frequency at different laser 
intensities with careful calculations of the intensity-dependent dipole 
shift will still allow for very accurate extrapolation to zero Stark
shift.

Collisional shifts, vanishingly small in the beam experiment, may
contribute as well in the lattice experiment. For fermions
this shift will be absent at the temperatures considered.

\section{Discussion and conclusion}
The beam and lattice experiment both promise signal strengths and
linewidths that should make a measurement with $1\,{\rm kHz}$
resolution possible. Improvements beyond this level also seem feasible.
Standard frequency comb technology
easily allows an absolute frequency measurement at this
accuracy. A fiber-laser based frequency comb
around $1.5\,\mu{\rm m}$ or a titanium-sapphire laser based
frequency comb (after frequency doubling) may be used for this purpose.

The main obstacles to be overcome are the residual Doppler
linewidth for the beam experiment and the dipole and collisional
shifts for the lattice experiment. Another factor to be considered
in both experiments is the Zeeman shift due to a stray magnetic
field. 
A solution is to measure
only an $\rm{M}=0 \rightarrow \rm{M}=0$ transition. For $^4$He, this can be
achieved by simply exciting with linearly polarised light.
For $^3$He, it will be important to shield
stray magnetic fields and measure both $-\frac{1}{2} \rightarrow
-\frac{1}{2}$ and $+\frac{1}{2} \rightarrow +\frac{1}{2}$ transitions and take
the average.

What will an absolute frequency measurement at or below the $1\,{\rm kHz}$
level in $^4$He (or $^3$He) test? In a recent paper Morton, Wu and
Drake \cite{Morton06a} have tabulated the most up-to-date
experimental and theoretical ionisation energies, both for $^4$He
and $^3$He. The theoretical uncertainties for the $2\,^3{\rm S}_1$
and $2\,^1{\rm S}_0$ states are larger than the experimental error for $^4$He for both states. 
Experiment and theory agree to well within the error bars.
It may be expected that the theoretical error in the 1.557 $\mu$m
\emph{transition} frequency will be smaller than the
quadratic sum of the errors in the ionisation energy of the
metastable states due to cancellation effects \cite{Pachucki06b}. Therefore, a
measurement of the $2\,^3{\rm S}_1$ - $2\,^1{\rm S}_0$ transition
frequency will not only provide a direct and accurate link between
the ortho (triplet) and para (singlet) helium system but will also
test QED calculations more stringently than the existing data.

The isotope shift can be calculated 
with very high accuracy. Using the most recent values for
the nuclear masses and the most recent evaluation of the difference in the square of the
nuclear charge radii, i.e., 1.0594(26) fm$^2$ \cite{Morton06b,Shiner95},
we deduce a theoretical transition isotope shift of 8034.3712(9) MHz.
A measurement of the isotope shift thus provides a very sensitive test
of theory. It may also be interpreted as a measurement of the
difference in nuclear charge radius of $^4$He and $^3$He. A
measurement at the 1 kHz level will provide this difference
with an accuracy of 0.001 fm~\cite{Drake05}, similar to the
accuracy obtained from isotope shift measurements on the 
$2\,^3{\rm S}_1$ - $2\,^3{\rm P}$ transition. These have been performed 
with $\sim$5 kHz
accuracy \cite{Morton06b,Shiner95}, limited by the 1.6 MHz natural linewidth of 
that transition.
The 8 Hz natural linewidth of the 1.557 $\mu$m transition and the 
possibilities to improve the beam and lattice experiment further
via sub-Doppler cooling resp. uncoupling of the trapping
and spectroscopy lasers, may push the accuracy below the 1 kHz level
providing new challenges to theorists calculating ionisation energies
and, in the case of the isotope shift, provide the most accurate data
on differences in nuclear charge radius and nuclear masses of the
isotopes involved.

\end{document}